\documentclass[5p,twocolumn]{elsarticle}

\usepackage{graphicx}
\usepackage{dcolumn}
\usepackage{pifont}
\usepackage{bm}
\usepackage{amsmath}
\usepackage{float}
\usepackage{txfonts}

\journal{Elsevier}

\begin{document}
\title{The maximum interbubble distance in relation to the radius of spherical stable nanobubble in liquid water: A molecular dynamics study}

\author{Song-Nam Hong}
\author{Song-Hyok Choe}
\author{Un-Gi Jong}
\author{Myong-San Pak}
\author{Chol-Jun Yu\corref{cor}}
\ead{ryongnam14@yahoo.com}

\cortext[cor]{Corresponding author}

\address{Chair of Computational Materials Design (CMD), Faculty of Materials Science, Kim Il Sung University, \\ Ryongnam-Dong, Taesong District, Pyongyang, Democratic People's Republic of Korea}

\begin{abstract}
The mechanism of superstability of nanobubbles in liquid confirmed by many experimental studies is still in debate since the classical diffusion predicts their lifetime on the order of a few microseconds. In this work, we study the requirement for bulk nanobubbles to be stable by using molecular dynamics simulations. Periodic cubic cells with different cell sizes and different initial radii are treated to simulate the nanobubble cluster, providing the equilibrium bubble radius and the interbubble distance. We find out that for nanobubble with a certain radius $R$ to be stable, the interbubble distance should be smaller than the maximum interbubble distance $L^*$ being proportional to $R^{4/3}$.
\end{abstract}


\maketitle

\section{\label{sec:intro}Introduction}
In the past decade, nanobubbles (NBs) in solution have been attracting the great attention due to their promising applications in various industrial fields such as medicine, agriculture, mining, environment, etc.~\cite{Demangeat,Seddon,Alheshibri} In fact, they are being widely used for molecular imaging and therapy in medicine~\cite{Wang,Hernandez,Oda,Brubler,Yang,Nguyen}, for promoting the growth of plants by enhancing the metabolism in agriculture, for increasing the flotation recovery of fine and ultrafine mineral particles in mining~\cite{Rahman,Peng}, and for treating the wastewater by utilizing those filled with various kinds of gas like ozone in environment~\cite{Terasaka,Rehman,Zheng}. This is associated with their unique mechanical and physicochemical characteristics such as extremely high specific area (surface area per volume), small buoyant force, free radical generation and high energy release in the gas-liquid interface before their collapse~\cite{Demangeat}. In spite of such successful applications, the main issue of the stability and size of NBs, which is important to understand the behaviour of water including NBs, is yet unclear.

In general, macrobubbles of over milimeters diameter have a short lifetime like that they rise quickly in liquid and burst at the surface due to a large buoyant force. Microbubbles also have tendency to readily collapse due to their gradual decrease in size and long stagnation. Being gas filled nanocavities with a diameter from 10 to 200 nm, however, NBs were found from many experimental studies~\cite{Ushikubo,Ohgaki,Duval,Liu1} to have an amazing logevity of hours, days, or even months. In fact, NBs behave as solid nanoparticles and have no tendency to rise in liquid due to a disappearance of buoyant force. Ushikubo {\it et al.}~\cite{Ushikubo} have reported that NBs filled with pure O$_2$ gas are stable for days, although those filled with air can survive for less than 1 h. They attributed such a long lifetime of NBs to the negatively charged gas-liquid interface, which in turn creates repulsion forces preventing their coalescence, and to the supersaturation of liquid, resulting in the reduce of the concentration gradient between the interface and liquid. Ohgaki {\it et al.}~\cite{Ohgaki} also have shown that the bulk NBs can exist for over 2 weeks and found out that these bubbles were packed closely each other. From these observations, they suggested a shielding mechanism between neighboring bubbles, which acts to keep them from dissolution when the interbubble distance is sufficiently small.

However, there is still lack of satisfactory explanation or consensus theory about such superstability of the bulk NBs. According to the theory of classical diffusion, the bulk NBs were expected to exist for only a few microseconds~\cite{Liu,Matsumoto}. In this theory, the internal pressure inside the bubble is given by $P =2\gamma/R$ due to Young-Laplace law, where $\gamma$ is the interfacial tension of the bubble wall and $R$ the bubble radius. For $R=50$ nm and $\gamma=72$ mN/m, the internal pressure was determined to be 3 MPa. Because of such extremely high internal pressure, the gas in the bubble should rapidly dissolve into the surrounding liquid in less than a few microseconds~\cite{Demangeat}. This is more obvious in the bulk NBs for its radius being much smaller than that of curvature in the surface NBs of the same size, and thus, the stability of bulk NBs is more problematic. Even NBs can not exist according to the classical thermodynamic theory.

Previously, many molecular dynamics (MD) studies on the properties of NBs were carried out. Most of MD simulation studies were dedicated to the existence of surface NBs, which are nanoscopic hemisphere found at the solid-liquid interface and observed earlier in 1990s~\cite{Brenner,Seddon1,Lohse,Zhu,Weijs2,Maheshwari}. On the other hand, the bulk NBs suggested much later in 2007~\cite{Jin} were studied by relatively small number of MD works~\cite{Weijs1,Matsumoto1,Lugli,Man}. Weijs {\it et al.}~\cite{Weijs1} have studied the nucleation, growth, and final size of bulk NBs through MD simulation of binary mixture of simple fluids, each representing the liquid and the gas respectively. 
They showed that the bulk NBs should be close enough to be stable by exhibiting that the bubble, which was stable with proper spacing, shrank with the doubled interbubble distance. Matsumoto~\cite{Matsumoto1} investigated the stability of bulk NBs of different $R$ with fixed interbubble distance, finding out that there exists a critical bubble radius for stable NB cluster, and successfully explained such behavior as well as collapse and stability of bubbles using pressure balance.

From both experimental and MD simulation studies, one can notice that the interbubble distance ($L$) is an important factor in stabilizing the bulk NBs and it is likely to be in a close relation with the bubble radius ($R$). In fact, the bulk NBs reported by Ohgaki {\it et al.}~\cite{Ohgaki} have an interbubble distance of $10R$ or less with $R\approx50$ nm. In another MD study by Weijs {\it et al.}~\cite{Weijs1}, the ratio of $L/R$ is about 7 at most when $L$ is fixed as 30 nm and $R\approx 4$ nm. Thus, the maximum interbubble distance $L^*$ seems not to be simply in a linear relation to $R$ over the whole nanoscopic range, and in order to clarify this relation a systematic study on NBs of different radii is required. Since it is cumbersome to estimate the maximum interbubble distance from experiment, the simulation method could be advisable. To the best of our knowledge, however, few systematic MD studies on this problem can be found.

In this paper, we aim to clarify the relation between the maximum interbubble distance and the bubble radius by performing MD simulations of NB clusters with different values of $L$ and $R$. We organize the paper as follows: In Section~\ref{sec_theoretics} computational details are outlined and the equation which Matsumoto~\cite{Matsumoto1} used for explanation of NB stability from a viewpoint of pressure balance is explained. Next, in Section~\ref{sec_result}, the results of the MD simulations are presented along with the comparison with the prediction of aforementioned equation as well as the results from other studies. Finally, the summary of findings in this study is given in Section\ref{sec_con}.

\section{\label{sec_theoretics}Method}
\subsection{Criterion for Stable Nanobubbles}
The stability of nanobubble (NB) can be explained by the pressure balance introduced in Ref.~\cite{Matsumoto1}. The pressure balance of a spherical NB with a radius of $R$ to surrounding liquid with a surface tension of $\gamma$ is often described by the Young-Laplace (Y-L) equation as follows,
\begin{equation}
\label{eq_p-liquid}
P_v=P_l+\frac{2\gamma}{R}
\end{equation}
where $P_v$ is the internal pressure of  the bubble and $P_l$ the pressure of surrounding liquid, respectively. It should be noted that in the Y-L equation~(\ref{eq_p-liquid}) the pressure difference $\Delta P=P_v-P_l$ diverged as the bubble size decreases like $R\rightarrow0$.

If the liquid is assumed to be water, the pressure of liquid water can be determined from the mass density using the equation of state (EOS). Here, the liquid density can be defined as the total mass of water molecules divided by the volume they occupy. As assuming the bubble to have a shape of sphere, the liquid density can be written as follows,
\begin{equation}
\label{eq_density}
\rho_l=\frac{mN}{V-\frac{4\pi}{3}R^3}
\end{equation}
where $m$ and $N$ are the mass and number of water molecules respectively, and $V$ the cell volume. For the EOS of liquid water, it was suggested that the water pressure should be a linear function of liquid density in the range of swelled state, in which the density is less than that of liquid water in normal state (1000 kg/m$^3$)~\cite{Matsumoto1}. Then, the pressure of liquid water ($P_l$) can be provided as follows, 
\begin{equation}
\label{eq_p-bulkWater}
P_l=A\rho_l+B=A\frac{mN}{V-\frac{4\pi}{3}R^3}+B
\end{equation}
where $A$ and $B$ are the parameters that will be determined during the linear fitting of simulation result, respectively.

From the above Eqs.~(\ref{eq_p-liquid}) and (\ref{eq_p-bulkWater}), the stability condition of the bulk NB can be derived as follows,
\begin{equation}
\label{eq_p-balance}
P_v-\frac{2\gamma}{R}=A\frac{mN}{V-\frac{4\pi}{3}R^3}+B
\end{equation}
By solving this equation~(\ref{eq_p-balance}) with the given values of $V$ and $N$ numerically and/or analytically, we can get two solutions for the bubble radius $R$, which correspond to the equilibrium radii of stable bubble (bigger $R$) and instable bubble (smaller $R$). Based on this equation, therefore, we can derive the relationship between the interbubble distance and the bubble radius for stable NBs by performing molecular dynamics (MD) simulations.

\begin{figure*}[!th]
\centering
\includegraphics[clip=true,scale=0.45]{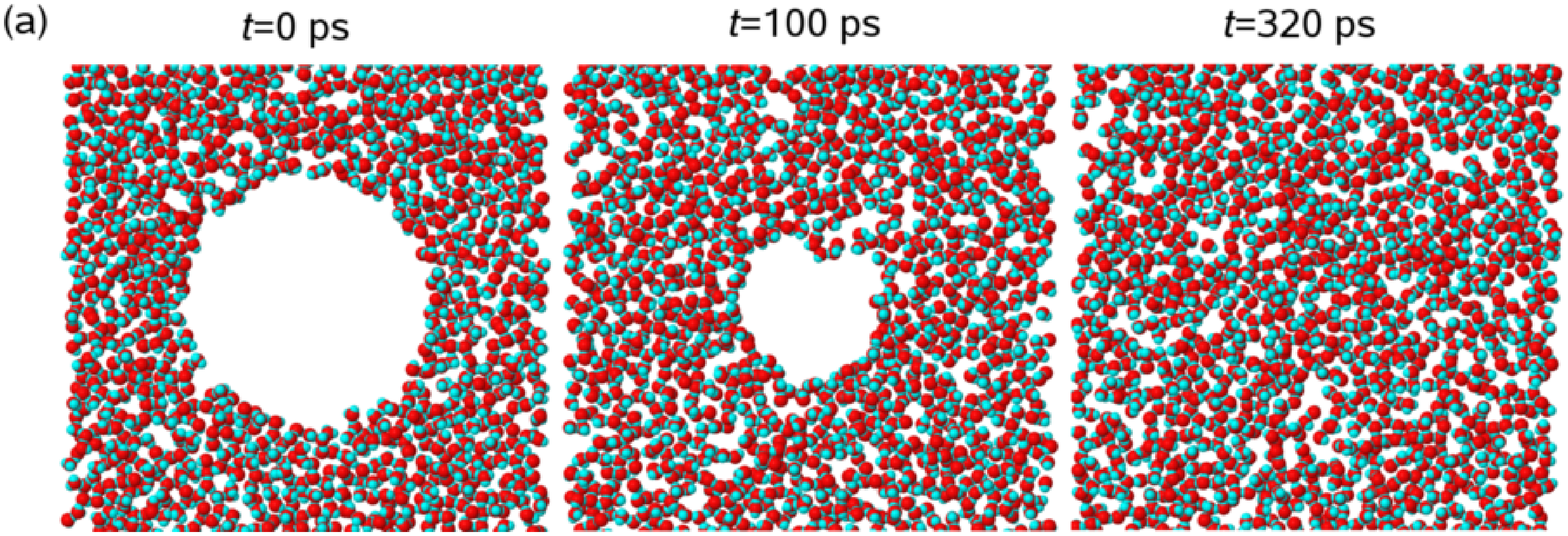} \\
\includegraphics[clip=true,scale=0.4]{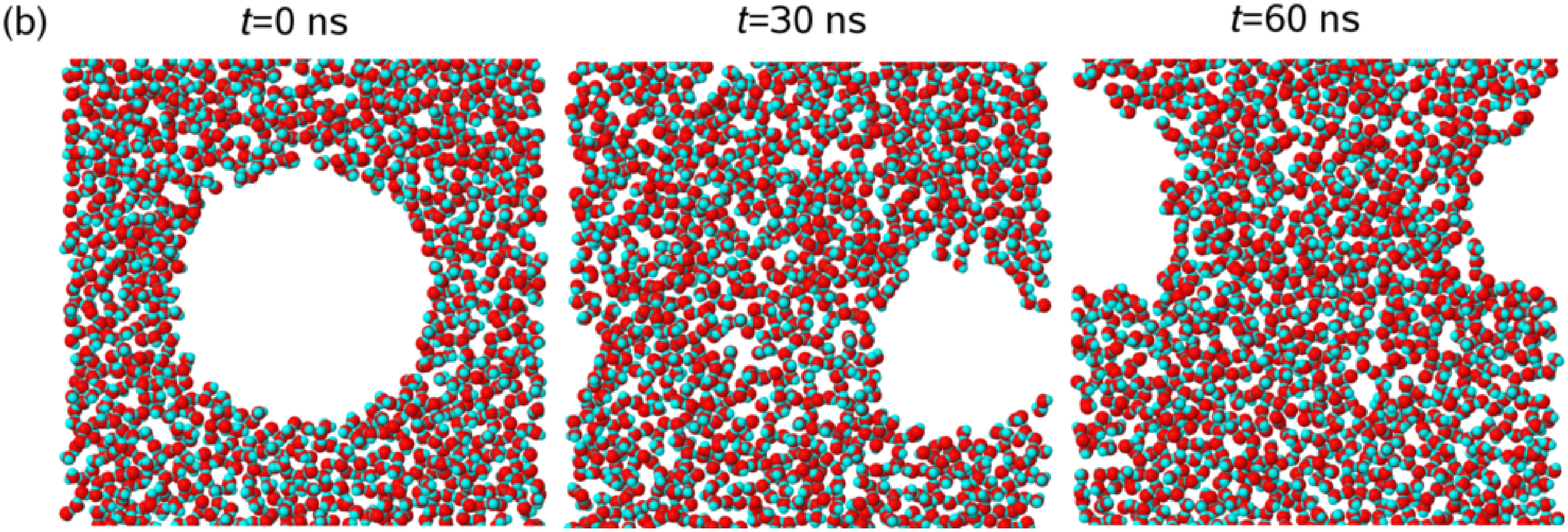} 
\caption{\label{fig_snapshot}Snapshots from simulations of nanobubble models of cell size $L=10$ nm with different initial radius of cavity: (a) $R_\text{ini}=2.6$ nm and (b)  $R_\text{ini}=2.7$ nm. Each snapshot is obtained from a slice with the thickness of 0.4 nm, which is taken near the center of a bubble.}
\end{figure*}

\subsection{Molecular Dynamics}
The MD simulations are carried out using the GROMACS program (version 4.52)~\cite{gromacs}. Liquid is simulated by a great number of water molecules and cavities are created by removing a group of water molecules inside to simulate the bulk NB. The interaction between water molecules is described by CHARMM27~\cite{charmm} force field with a cutoff radius of 1.2 nm. The canonical ensemble ($N$, $V$, $T$ constant) MD simulations are performed to get the equilibrium state at the given temperature ($T$), where the velocity rescaling thermostating procedure is employed to keep the temperature constant. The initial velocities of atoms are generated according to the Maxwell distribution at room temperature of 300 K.

Firstly, a periodic cubic cell with a certain length ($L$), which is filled with a certain number of water molecules determined by the mass density of $\sim$ 1,000 kg/m$^3$, are equilibrated at 300 K for the total simulation time of 50 ps with a tiemestep of 2 fs. Then, some water molecules in the spherical region of radius ($R$) are removed from the cell to simulate the bulk NB. Due to the periodic boundary condition (PBC) in all the three directions to model the nanobubble cluster, the cell size $L$ directly serves as the interbubble distance. Subsequently, the NVT simulation is performed on this NB containing cell for the total time of 10 ns to verify whether the bubble is stable or not. If the bubble is proved to be stable, the simulation is continued up to 60 ns to examine the change of the bubble radius over simulation time. The bubble radius is estimated from its spherical volume determined by counting the number of vacant minicells with the dimension of $0.05\times0.05\times0.05$ nm composing the cell.

In order to find out the relation between the maximum interbubble distance $L^*$ and the bubble radius $R$, above procedure is repeated for the cells with different cell sizes and different radii of bubble. The cell lengths taken in this work are 5, 8, 10, 12, and 15 nm, and in each case of $L$, the bubble radius $R$ increases systematically from a sufficiently small value in order to determine the least stable bubble radius $R^*$. 

\section{\label{sec_result}Results and Discussion}
First, we performed NVT simulations of cubic cells filled by a determined number of water molecules with different cell lengths ($L=5, 8, 10, 12, 15$ nm) to obtain the equilibrated bulk liquid models at room temperature of 300 K. The total time was 50 ps, by which the systems were confirmed to be fully equilibrated. Then, cavities with different initial radii ($R_\text{ini}$) were created by removing the corresponding water molecules in each cubic cell. Subsequently, the NB-containing cubic cells were equilibrated by performing NVT simulations for the total time of 10 ns. It was found that according to the cell length and initial cavity radius some NBs collapsed in a relatively short time within hundreds of picoseconds, while some NBs could exist for a long time of over 60 ns.

Figure~\ref{tbl_result} shows some snapshots of NVT simulations for the NB containing cubic cells of $L=10$ nm with two different initial cavity radii of $R_\text{ini}=2.6, 2.7$ nm.   As shown in Figure~\ref{fig_snapshot}(a), the bubble with $R_\text{ini}=2.6$ nm was found to shrink and completely disappear within 320 ps. For the case of $R_\text{ini}=2.7$ nm shown in Figure~\ref{fig_snapshot}(b), however, the bubble was found to contract to the equilibrium size of $R_\text{equ}\approx2.1$ nm almost instantly (within 100 ps) and remain in its stable state for over 60 ns. It is worth noting that once the NB is found to be stable, it has a fluidity in liquid similar to the Brownian movement of nanoparticle in liquid; its center is not fixed at the center of the cell but moves around after it has become equilibrium. 

\begin{figure}[!b]
\centering
\includegraphics[clip=true,scale=0.52]{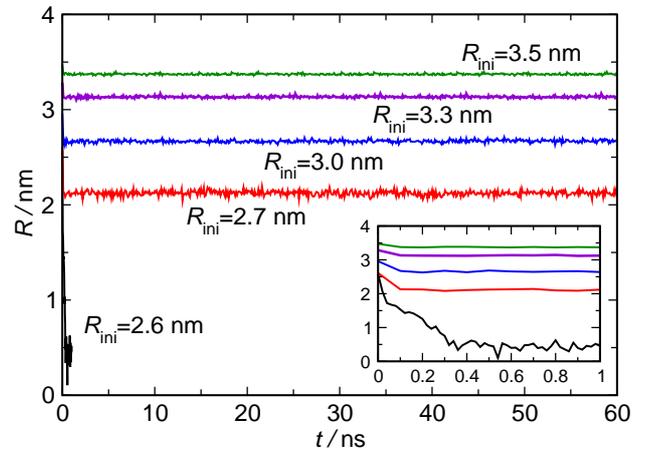}
\caption{\label{fig_r-t}Variation of bubble radius in cubic cell of cell length $L=10$ nm with different initial cavity radii ($R_\text{ini}$) as the simulation time advances. Inset shows details in the time period from 0 to 1 ns.}
\end{figure}
In Figure \ref{fig_r-t} we present the variation of the bubble radius in the NB containing cubic cell of the cell length $L=10$ nm with different initial cavity radii of $R_\text{ini}=2.6, 2.7, 3.0, 3.3, 3.5$ nm. As already above mentioned, the bubble with the initial cavity radius of $R_\text{ini}=2.6$ nm is evident to collapse due to an abrupt decrease of the cavity radius to almost zero within 320 ps. On the other hand, for the rest stable nanobubbles, the initial cavity radius was seen to quickly arrive at the equilibrium value within 100 ps and after that do an equilibration around that value. It should be noted that the shift of cavity radius from the initial value to the equilibrium value becomes more significant for smaller $R_\text{ini}$. 

\begin{table}[t]
\small
\caption{\label{tbl_result}Result of MD simulations for NB containing cubic cells with cell length ($L$), initial cavity radius ($R_\text{ini}$) and certain number of water molecules ($N$), where $t_\text{c}$ is the time of bubble collapse and $R_\text{equ}$ is the average radius of stable bubble in its equilibrium state.}
\begin{tabular}{cccccc}
\hline
$L$ (nm) & $R_\text{ini}$ (nm) & $N$ & Stable & $t_{c}$ (ps) & $R_\text{equ}$ (nm) \\
\hline
5   & 1.5 & 3672     & no  & 18     & $-$ \\
5   & 1.6 & 3575     & yes & $-$   & 1.33 \\
8   & 2.2 & 15645   & no  & 10     & $-$ \\
8   & 2.3 & 15440   & yes & $-$   & 1.75 \\
10 & 2.4 & 31305   & no  & 54     & $-$ \\
10 & 2.6 & 30774   & no  & 320   & $-$ \\
10 & 2.7 & 30479   & yes & $-$   & 2.12 \\
10 & 3.0 & 29434   & yes & $-$   & 2.71 \\
10 & 3.3 & 28194   & yes & $-$   & 3.13 \\
10 & 3.5 & 27235   & yes & $-$   & 3.37 \\
12 & 3.1 & 53549   & no  & 714   & $-$ \\
12 & 3.2 & 53119   & yes & $-$   & 2.43 \\
15 & 2.7 & 107870 & no  & 26     & $-$ \\
15 & 3.4 & 105624 & no  & 236   & $-$ \\
15 & 3.5 & 104669 & yes & $-$   & 2.79 \\
\hline
\end{tabular} 
\end{table}
Similar results were obtained for other cells with different cell lengths. Table~\ref{tbl_result} present the main result of simulations for NB containing cubic cells with different cell length $L$, different initial cavity radii $R_\text{ini}$, and the corresponding number of water molecules $N$. The bubbles could be divided into two types: to be unstable, for which the collapse time ($t_\text{c}$) is given, and to be stable, for which the equilibrium radius ($R_\text{equ}$) is presented. For instance, in the case of cell with the cell length $L=10$ nm, the bubbles with the initial cavity radii $R_\text{ini}=2.4, 2.6$ nm were found to be unstable, whose the collapse times were determined to be 54 and 320 ps, respectively. Thus, it can be said that smaller bubble collapses more quickly. On the contrary, the bubbles with the initial cavity radii $R_\text{ini}=2.7, 3.0, 3.3, 3.5$ nm turned out to be stable, whose equilibrium radii were estimated to be $R_\text{equ}=2.12, 2.71, 3.13, 3.37$ nm, respectively. 

It should be noted that although the stable bubble moves constantly and changes in size, the interbubble distance is kept constant as $L$ due to the periodic boundary condition. Therefore, we can derive the relation between $L$ and $R$ based on the results given in Table~\ref{tbl_result}. To this end, the critical initial radius for stable bubble should be determined for all the cells. For the case of $L=10$ nm, it was determined to be $R^*_\text{ini}=2.7$ nm, and the corresponding critical equilibrium radius was $R^*_\text{equ}=2.12$ nm. Such consideration is carried out for the cases of different cell lengths $L=5, 8, 10, 12, 15$ nm, resulting in $R^*_\text{equ}=1.33, 1.75, 2.12, 2.43, 2.79$ nm. Meanwhile, it was found that the bubble with $R_\text{ini}=2.7$ nm is stable in the simulation cell of $L=10$ nm, but it collapses within 26 ps in the cell of $L=15$ nm. This indicates that wider spacing requires larger radius for stable NBs; longer interbubble distance prefers to larger radius of stable bubble, as already pointed out in the previous works~\cite{Weijs1,Bunkin1,Bunkin2}.

\begin{figure}[!t]
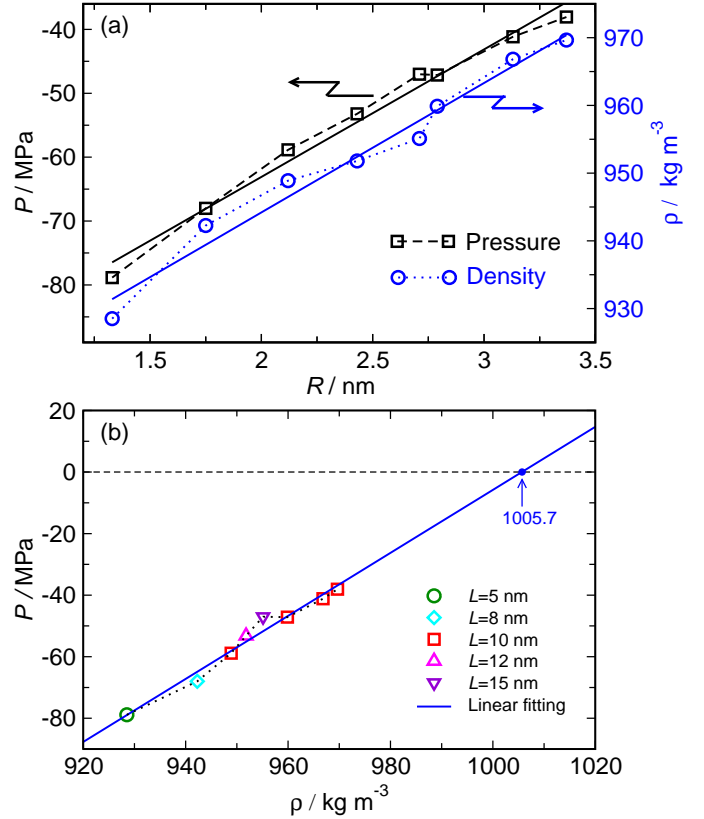

\includegraphics[clip=true,scale=0.52]{fig3a.eps} \\
\includegraphics[clip=true,scale=0.52]{fig3b.eps}
\caption{\label{fig_rho-p}(a) Liquid water pressure ($P$) and density ($\rho$) as the equilibrium radius of bubble ($R$) increases, and (b) liquid water pressure as the density increases, where the pressure becomes positive over $\rho=1005.7$ kg/m$^3$. Linear fitting lines are given by interpolation and extrapolation.}
\end{figure}
\begin{figure}[!t]
\centering
\includegraphics[clip=true,scale=0.52]{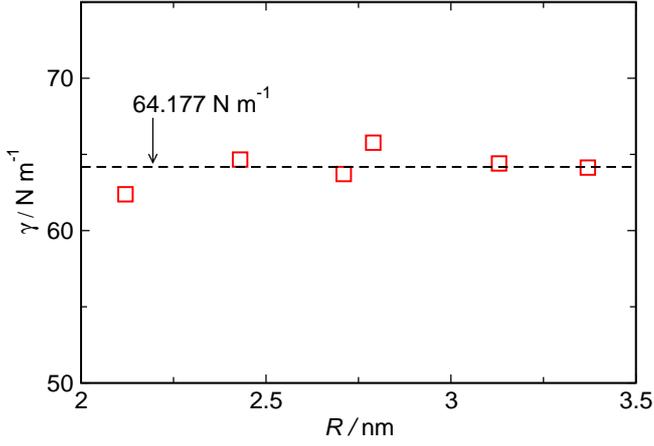}
\caption{\label{fig_gamma}Surface tension as the bubble radius increases, with the average value of 64.177 N/m.}
\end{figure}

Now, we can identify this relationship analytically using the pressure balance given in Eq.~\ref{eq_p-balance}. In this equation, the internal pressure of the bubble should be zero, because the bubble does not contain any water molecule as can be seen in Figure~\ref{fig_snapshot}, {\it i.e.}, $P_v=0$. On the other hand, the pressure of surrounding liquid water was obtained to be a large negative value, indicating that the liquid water was in a swelled state. This was verified by evaluating the density of liquid water using Eq.~\ref{eq_density}, being smaller than the normal liquid water density of 1000 kg/m$^3$. In Figure~\ref{fig_rho-p}(a) we show the obtained pressure and density of surrounding liquid water as increasing the equilibrium radius of bubble, which were fitted into linear functions of bubble radius like $P(R)=19.960R-103.015$ (MPa) and $\rho(R)=19.111R+905.973$ (kg/m$^3$). Figure~\ref{fig_rho-p}(b) shows the pressure as a linear function of density as suggested in Eq.~\ref{eq_p-bulkWater} for the EOS of liquid water. Fitting the result yielded a linear function of $P(\rho)=1.024\rho-1029.430$ (MPa). We extrapolated this function from 920 to 1020 kg/m$^3$, and then the pressure becomes positive over the density of 1005.7 kg/m$^3$ being very close to the density of water at normal state, as shown in Figure~\ref{fig_rho-p}(b). This implies that the linear function is valid in the density region of normal state as well as swelled state. Thus, the obtained fitting parameters ($A=1.024$, $B=-1029.430$) can be used in safe to identify the relationship between the interbubble distance and the bubble radius. To this end, we rewrite Eq.~\ref{eq_p-balance} using the cell length $L$ (interbubble distance) instead of volume $V$ as follows,
\begin{equation}
\label{eq_p-bal-r-l}
-\frac{2\gamma}{R}=A\frac{mN}{L^3-\frac{4\pi}{3}R^3}+B
\end{equation}
where the surface tension $\gamma$ can be obtained from the liquid water pressure and the bubble radius using Eq.~\ref{eq_p-liquid}. Figure~\ref{fig_gamma} shows the surface tension as the bubble radius increases, indicating a weak fluctuation around the average value of 64.177 N/m.

\begin{figure}[!t]
\centering
\includegraphics[clip=true,scale=0.52]{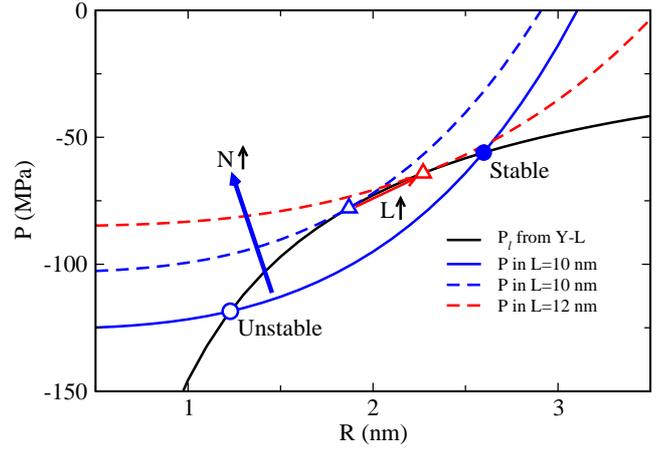}
\caption{\label{fig_balance}Behavior of stable bubble radius from pressure balance. Black solid line presents the liquid pressure $P_l$ derived from Young-Laplace equation, blue solid and dashed lines the pressure from the fitted function in the cubic cell of $L=10$ nm, and red dashed line in $L=12$ nm.}
\end{figure}

For an interpretation of Eq.~\ref{eq_p-bal-r-l}, we plots the liquid pressure $P_l$ derived from Y-L equation (Eq.~\ref{eq_p-liquid}) and those from the fitted linear function with the parameters $A$ and $B$ as functions of bubble radius in Figure~\ref{fig_balance}. Black solid line presents the $P_l$ curve from the Y-L equation and the colored lines present the curves from fitted functions. The curves of pressure from the fitted function shift upward when increasing the number of water molecules $N$, and this will contact with the $P_l$ curve at the certain value of $N$ (see blue solid and dashed lines in Figure \ref{fig_balance}). This point of contact accounts for the least stable bubble radius, {\it i.e.}, the critical equilibrium radius ($R^*_{equ}$) for a given interbubble distance $L$, and vice versa, it is the maximum interbubble distance ($L^*$) for a given bubble radius $R$. Meanwhile, this point moves towards the right along the $P_l$ curve when increasing the cell length of $L$ (from blue dashed line to red dashed line in Figure \ref{fig_balance}). Therefore, the relation between $L^*$ and $R$, the major concern in this study, can be obtained as follows (see Appendix for details),
\begin{equation}
\label{eq_l-r}
L^*=\left(-\frac{2\pi}{3\gamma}R^4\left(\frac{4\gamma}{R}+3B\right)\right)^{1/3}
\end{equation}
where the first term $4\gamma/R=-2P_l$, which is approximately 300 MPa when $R=1$ nm, and the second term $3B\approx$ 3000 MPa. That is, the term ${4\gamma}/R$ is much smaller than the term $3B$ when $R$ increases; it is negligible for $R>10$ nm. Therefore, Eq.~\ref{eq_l-r} can be approximated as follows
\begin{equation}
\label{eq_appr-result}
L^*\approx\left(-\frac{2\pi{B}}{\gamma}\right)^{1/3}R^{4/3}=\left(-\frac{4\pi{B}}{P_l}\right)^{1/3}R
\end{equation}
This was proved to be a good approximation since the curve derived from this equation (Eq.~\ref{eq_appr-result}) was almost identical to the curve derived from Eq.~\ref{eq_l-r}.
%

\begin{figure}[!b]
\centering
\includegraphics[clip=true,scale=0.52]{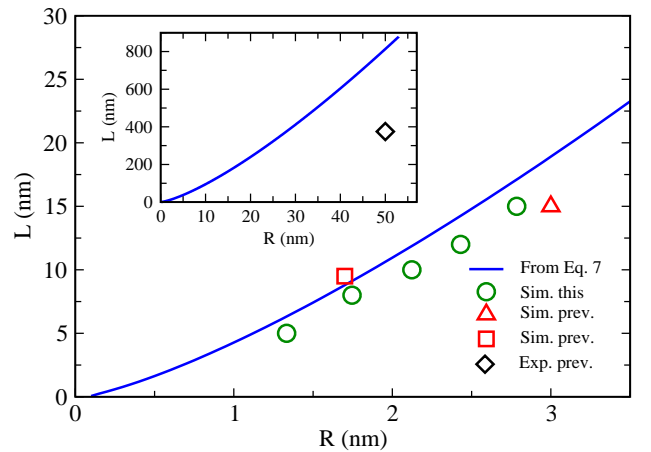}
\caption{\label{fig_l-r}Interbubble distance versus bubble radius. Sim, exp and prev stand for simulation, experiment and previous result, respectively. Green circles represent the equilibrium bubble radii $R^*_\text{equ}$ from this work, red triangle from Ref.~\cite{Weijs1}, red square from Ref.~\cite{Matsumoto1}, and black diamond from Ref.~\cite{Ohgaki}. Inset shows the curve extended up to $R=50$ nm.}
\end{figure}
\begin{figure}[!t]
\centering
\includegraphics[clip=true,scale=0.52]{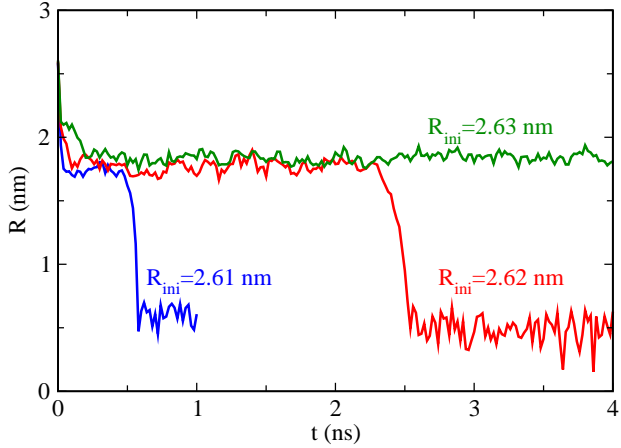}
\caption{\label{fig_r-t-1}Bubble radius as the simulation time advances, with $R_\text{ini}=2.61, 2.62, 2.63$ nm.}
\end{figure}

We verified a validity of this approximate equation by comparing with those from simulation and experiment. Figure~\ref{fig_l-r} shows the interbubble distance as a function of bubble radius obtained from Eq.~\ref{eq_appr-result} together with some MD simulation results and available experimental result. It was revealed that the data of $R^*_\text{equ}$ and corresponding $L$ from our MD simulation (Table \ref{tbl_result}) represented as green circles were very close to the prediction of Eq. \ref{eq_appr-result} (blue line) with slightly being larger. Such difference can be attributed to the design of simulation. Since we determined $R^*_\text{equ}$ by systematically increasing $R_\text{ini}$ with the interval of 0.1 nm, it was close to the least value of stable bubble radius but not the same; it must be larger than the prediction. It might be thought that more deliberate consideration with smaller interval of $R_\text{ini}$ can give closer result. In fact, as shown in Figure \ref{fig_r-t-1}, we performed such simulation with the cell length of $L=10 $ nm with the interval of 0.01 nm, resulting in $R^*_\text{equ}=1.864 $ nm that is almost indentical to the prediction of 1.869 nm. Moreover, the bubble radius $R$ for the point of contact in Figure \ref{fig_balance} was slightly underestimated for $R^*_\text{equ}$ since it was not the stable equilibrium point. As mentioned earlier, even if the bubble would be stable, its radius $R$ constantly fluctuated around the equilibrium value $R_\text{equ}$ over time and the range of its change expanded as $R_\text{equ}$ approached to $R^*_\text{equ}$. 

And once $R$ gets smaller than the unstable threshold (open circle in Figure \ref{fig_balance}), the bubble shrinks and disappears~\cite{Matsumoto1} as being apparent in Figure \ref{fig_r-t-1}. In this figure, the bubble with $R_\text{ini}=2.62$ nm was found to survive for 2.5 ns before it suddenly collapsed. For NBs to prevent such collapse and to become stable, $R^*_\text{equ}$ should be estimated to be slightly larger than $R$ for the point of contact to ensure that its change does not exceed the unstable threshold. Therefore, our simulation result was in good agreement with the curve of $L^*$. On the other hand, the MD simulation results from Matusmoto~\cite{Matsumoto1} and Weijs {\it et al.}~\cite{Weijs1} (red square and triangle in Figure \ref{fig_l-r}) were found to agree with our result as well. The experimaental result from Ohgaki~\cite{Ohgaki} was also shown in the inset as a black diamond. These points are near or under the curve of $L^*$, meaning that the interbubble distances are small sufficiently; it is less than the maximum interbuble distance $L^*$, hence support for Eq.~\ref{eq_appr-result}.

Consequently, this relationship between the maximum interbubble distance and the bubble radius $-$ derived from the pressure balance, is not only valid when $R$ is in the range of a few nanometers considered in this work, but also expected to be applicable in the cases of larger $R$ and $L$, where MD simulation is currently not affordable becasue of immense calculation cost. For NBs to be stable, therefore, the interebubble distance should be smaller than $L^*$, which is proportional to $R^{4/3}$.

\section{\label{sec_con}Conclusions}
In this work, we have performed molecular dynamics simulations to investigate the stability of nanobubbles in liquid water. Periodic cubic cells with different cell sizes and different initial bubble radii were treated to simulate the nanobubble cluster with a certain interbubble distance. The relationship between the interbubble distance and the bubble radius in stable nanobubble cluster was clarified. From both the MD simulations and the consideration from a viewpoint of pressure balance, it was revealed that for nanobubbles of a certain radius $R$ there exists the upper limit for interbubble distance. We have derived this limit, {\it i.e.} the maximum interbubble distance ($L^*$), to be proportional to $R^{4/3}$, and confirmed its validity through a comparison with the MD simulation results and available experimental result. We believe that this relationship is useful in practical applications of nanobubble by deteriming the stable bubble concentration in solution.

\section*{\label{ack}Acknowledgments}
This work was supported partially by the State Committee of Science and Technology, Democratic People's Republic of Korea, under the state research project ``Design of Innovative Functional Materials for Energy and Environmental Application'' (No. 2016-20). The calculations have been carried out on the HP Blade System C7000 (HP BL460c) that is owned by Faculty of Materials Science, Kim Il Sung University.

\section*{\label{note}Notes}
The authors declare no competing financial interest.

\appendix
\section*{\label{append}Appendix}
The relation between the simulation cell length and the stable bubble radius is clarified. The stability condition of nanobubble can be written as follows,
\begin{equation}
\label{ap1}
-\frac{2\gamma}{R}=A\rho+B \tag{A1}
\end{equation}
From this equation, the liquid density is given as follows,
\begin{equation} \label{ap2}
\rho=\frac{-\frac{2\gamma}{R}-B}{A} \tag{A2}
\end{equation}
Differentiating both side of the equation gives the following equation,
\begin{equation}
\label{ap3}
d\rho=\frac{2\gamma}{AR^2}dR \tag{A3}
\end{equation}
If the two solutions of Eq.~\ref{ap1} are assumed to be $R_1$ and $R_2$, the water densities at those points are as follows,
\begin{equation}
\label{ap4}
\rho_1=\frac{mN}{V-\frac{4\pi}{3}R_1^3},~~
\rho_2=\frac{mN}{V-\frac{4\pi}{3}R_2^3} \tag{A4}
\end{equation}
\begin{equation}
\label{ap5}
\rho_1V=mN+\frac{4}{3}\pi\rho_1R_1^3,~~
\rho_2V=mN+\frac{4}{3}\pi\rho_2R_2^3 \tag{A5}
\end{equation}
If subtracting Eq.~\ref{ap5} by sides, the following is obtained,
\begin{equation}
\label{ap6}
(\rho_1-\rho_2)V=\frac{4\pi}{3}\left(\rho_1R_1^3-\rho_2R_2^3\right) \tag{A6}
\end{equation}
If defining $\Delta\rho=\rho_1-\rho_2$, the above equation becomes
\begin{equation}
\label{ap7}
\Delta\rho V=\frac{4\pi}{3}\left(\rho_1R_1^3-\rho_1R_2^3+\Delta\rho R_2^3\right) \tag{A7}
\end{equation}
\begin{equation}
\label{ap8}
=\frac{4\pi}{3}\left[\rho_1(R_1-R_2)\left(R_1^2+R_1R_2+R_2^2\right)+\Delta\rho R_2^3\right] \tag{A8}
\end{equation}
If defining $\Delta R=R_1-R_2$ and considering $R_1\approx R_2$ close to the contact point, the following is obtained,
\begin{equation}
\label{ap9}
\Delta\rho V=\frac{4\pi}{3}\left(3\rho_1 \Delta R R_1^2+\Delta \rho R_1^3\right) \tag{A9}
\end{equation}
Considering Eq.~\ref{ap3} gives the following,
\begin{equation}
\label{ap10}
\frac{2\gamma}{AR_1^2}V=\frac{4\pi}{3}\left(2\rho_1 -\frac{B}{A}\right) \tag{A10}
\end{equation}
Considering Eq.~\ref{ap2} and $V=L^3$, the final equation can be obtained,
\begin{equation}
\label{ap11}
L=\left(-\frac{2\pi}{3\gamma}R_1^4\left(\frac{4\gamma}{R_1}+3B\right)\right)^{1/3} \tag{A11}
\end{equation}

\bibliographystyle{elsarticle-num-names}
\bibliography{Reference}

\end{document}